# Simultaneous generation and transfer of mechanical noise squeezing

Mungyeong Jeong[1], Hyojun Seok[2], Young-Sik Ra[3] & Junho Suh[1*]

[1]*Department of Physics, Pohang University of Science and Technology (POSTECH), Pohang, Korea*

[2]*Department of Physics Education, Seoul National University, Seoul, Korea*

[3] *Department of Physics, Korea Advanced Institute of Science and Technology, Daejeon, Korea*

[*]Corresponding author. Email: junhosuh@postech.ac.kr

**Abstract:** Optomechanical interactions between mechanical oscillators and an electromagnetic field induce controllable modifications in mechanical fluctuation. When multiple mechanical oscillators are coupled to a single electromagnetic mode, these interactions can be extended to utilize the electromagnetic mode as a mediator for distributing noise squeezing among different mechanical oscillators. We investigate the transfer of mechanical noise squeezing between two mechanical modes, enabled by a single microwave cavity mode which is strongly coupled to both mechanical modes. Noise squeezing in one mechanical mode (control) is achieved through parametric modulation of its resonance frequency via the optical spring effect. Simultaneously, optomechanical beam-splitter interaction is applied between the mechanical modes to transfer noise squeezing from the control mode to the other mode (target). Strong correlations between the quadratures of the two mechanical modes confirm that the observed squeezing in the target mode originates from the squeezing in the control mode. Remarkably, the observed squeezing transfer manifests noise characteristics of both single-mode and two-mode squeezing. This unique feature suggests that the squeezing transfer holds significant potential for enhancing precision measurements.

Tunable interactions in coupled oscillator systems have become increasingly important for generating diverse oscillator states and exploring novel coupled dynamics in optomechanical platforms. This tunability via amplitude or phase modulation of the coupling, both static and dynamic, enables precise tailoring of inter-mode dynamics as demonstrated in previous experiments of back-action evading[1,2], squeezing[3–8], entanglement[9–11], and coherent state transfer involving optical and mechanical modes[12,13]. Especially, coherent state transfer enables creating quantum states by transferring a state prepared in one oscillator to another thus places itself as an essential technique for quantum communication[14,15]. For instance, squeezed states can be transferred from various quantum systems to a mechanical mode[16–19], and in particular, it was proposed as an effective strategy to induce squeezing in magnons, which are challenging to generate otherwise[20–23]. Moreover, such coherent transfer generates correlations between the two oscillators, potentially leading to entanglement at the quantum level[11,13].

In this work, we demonstrate the transfer of noise squeezing between two mechanical modes through strong optomechanical interactions with single microwave cavity mode. We squeeze the motional fluctuation in one mechanical mode (control) via degenerate parametric amplification based on optical spring, and simultaneously apply a swap interaction to transfer the squeezing effect to the other mode (target). Correlations between the quadratures of the two mechanical modes confirm that the noise squeezing in the target mode originates from the control mode. We actively controlled the phase of the swap interaction to tune the correlation between the two mechanical modes, confirming a good agreement to theoretical predictions. At particular modulation phases, two-mode squeezing is identified in the quadrature correlations, revealing a unique feature of our technique to combine single-mode and two-mode squeezing in one experimental setup.

**Results**

**Microwave optomechanical system for cavity-mediated coupling of two mechanical modes**

Our device consists of a three-dimensional microwave cavity capacitively coupled to a $Si_3N_4$ membrane. The microwave cavity is machined from aluminum to form a reentrant cavity structure, as illustrated in **Fig. 1a**. Reentrant cavities provide efficient optomechanical couplings to produce motional sidebands even at ambient temperature[24]. The high-stress $Si_3N_4$ membrane (2 mm × 2 mm × 50 nm) is coated with a 30-nm-thick aluminum to enhance capacitive coupling to the cavity. This rectangular membrane supports multiple mechanical eigenmodes and these modes are the mechanical oscillators in our microwave cavity optomechanical system. The assembled device is placed in a vacuum chamber maintained at pressures below $10^{-3}$ mbar to minimize gas damping of membrane modes, and all experiments are performed at room temperature. We examine these multiple mechanical modes using optomechanically induced transparency (OMIT)[25,26], and identify two mechanical modes with significant optomechanical couplings ("control" and "target") as the mechanical oscillators in following experiments (**Fig. 1c**).

The cavity bottom is compliant (thickness ≈ 0.1 mm) and an external actuator is attached to adjust the vacuum gap between the membrane and the cavity tip ($D$)[27–29]. We note that this feature places tunability in cavity resonant frequency $\omega_c$ and single-photon optomechanical coupling rate $g_{0j}$ ($j$=1 for control and 2 for target). In the following experiments, the resonant frequency of the cavity is 3.50 GHz (= $\omega_c/2\pi$), and its dissipation rate is 38.9 MHz (= $\kappa/2\pi$). The control and target mechanical mode have resonant frequencies of 159.5 kHz (= $\Omega_1/2\pi$) and 351.1 kHz (= $\Omega_2/2\pi$), and their dissipation rates are 39.9 Hz (= $\gamma_1/2\pi$) and 13.8 Hz (= $\gamma_2/2\pi$), respectively. When a strong external drive ("pump") at frequency $\omega_d$ near cavity resonance is applied to raise intracavity photon number to $n_d$, the effective optomechanical coupling rate scales as $g_{0j}\sqrt{n_d}$ [30]. Three-dimensional cavities support a large number of pump photons (up to ~$10^{13}$ in our setup) boosting the effective coupling significantly[29,31–33], and the cooperativity $C_j = \frac{4g_{0j}^2 n_d}{\gamma_j \kappa}$ can exceed unity for both mechanical modes in our experiments. For instance, at $n_d = 2.88 \times 10^{13}$, we achieve $C_1 = 14.9$ and $C_2 = 1.90$.

The generation and transfer of mechanical squeezing is accomplished by modulating the pump on two frequencies, $2\Omega_1$ and $\Delta\Omega$ (= $\Omega_2 - \Omega_1$) as,

$$n_d(t) = \langle n_d \rangle [1 + \beta_s \cos(2\Omega_1 t + \phi_s) + \beta_t \cos(\Delta\Omega\, t + \phi_t)], \text{(Eq.2)}$$

where $\beta$ and $\phi$ are the amplitude modulation depths and phases for generation and transfer, denoted with subscripts "$s$" and "$t$", respectively. Our cavity optomechanical system operates in the unresolved sideband regime ($\kappa \gg \Omega_i$), where cavity photons dissipate faster than the mechanical oscillation periods. As a result, the time modulation of pump photon number induces a time-varying optomechanical backaction[34]. Under the condition of large detuning ($|\omega_d - \omega_c| \gg \Omega_j$) the system driven by $n_d(t)$ can be described by an effective Hamiltonian $H_{eff}$, which includes only interactions between two mechanical modes[34,35] (**Fig. 1d**):

$$\hat{H}_{eff} = \hbar \left[ \frac{\beta_t \sqrt{\delta\Omega_{opt,1} \delta\Omega_{opt,2}}}{2} \left( \hat{b}_1^+ \hat{b}_2 e^{i\phi_t} + \hat{b}_2^+ \hat{b}_1 e^{-i\phi_t} \right) + \frac{\beta_s \delta\Omega_{opt,1}}{4} \left( \hat{b}_1^+ \hat{b}_1^+ e^{i\phi_s} + \hat{b}_1 \hat{b}_1 e^{-i\phi_s} \right) \right]. \quad \text{(Eq.3)}$$

The first term represents the beam-splitter or transfer interaction between modes 1 and 2, and the second term provides single-mode squeezing for the control mode. In this effective Hamiltonian, the system can be regarded as a coupled system of two high-Q mechanical modes. Here, $\delta\Omega_{opt,j}$ represents the mechanical resonance shift of mode $j$ from the optical spring effect induced by the detuned microwave drive.

**Noise squeezing of mechanical mode via degenerate parametric amplification**

Degenerate parametric amplification of a mechanical oscillator is achieved by modulating its effective spring constant at twice of its own resonant frequency[3,4,8,34,36,37]. The gain in the degenerate parametric amplification is phase-dependent, resulting in negative gain, or de-amplification, on the quadrature orthogonal to the amplified one. When the mechanical oscillator is driven by random noise force, this de-amplified quadrature of motion results in reducing, or squeezing the fluctuations of mechanical motion.

We realize degenerate parametric amplification by modulating the effective spring constant of a mechanical oscillator through time-varying radiation pressure. This technique has been previously employed in cavity optomechanics experiments with optical light[4,34,37], and it eliminates the need for additional electrodes to apply a static voltage[8] in microwave cavity optomechanical systems. The pump tone at $\omega_d$ is applied with detuning from cavity $\Delta (\equiv \omega_d - \omega_c) = -\frac{\kappa}{2\sqrt{3}}$, and its amplitude is modulated at $2\Omega_1$. Here, the mechanical motion of control mode in the classical limit is described by the equation of motion:

$$m_1 \frac{d^2 x_1}{dt^2} + m_1 \gamma_1 \frac{dx_1}{dt} + \left( m_1 \Omega_1^2 + \delta k_{opt,1} \sin(2\Omega_1 t - \phi_s) \right) x_1 = F_1 \cos(\Omega_1 t + \phi_1), \quad \text{(Eq.4)}$$

where $m_j$ is the effective mass of the mechanical mode, $x_j$ is the displacement, $\delta k_{opt,j}$ is the optically induced modulation of the spring constant, and $F_j(t)$ is the external force for each mode $j$.

We measure the fluctuation of mechanical motion via the optomechanical sidebands generated by the microwave drive tone. The mechanical motion is expressed as $x_j(t) = X_j(t)\cos\Omega_j t + Y_j(t)\sin\Omega_j t$, with the noise amplitude given by $|x_j| = \sqrt{X_j^2 + Y_j^2}$. The noise amplitudes of control mode when parametric modulation is active ("on") or inactive ("off") satisfiy[36]:

$$|x_1|_{on} = |x_1|_{off} \left[\frac{\cos^2(\phi_1+\phi_s)}{\left(1+\frac{Q_1\delta k_{opt,1}}{m_1\Omega_1^2}\right)^2} + \frac{\sin^2(\phi_1+\phi_s)}{\left(1-\frac{Q_1\delta k_{opt,1}}{m_1\Omega_1^2}\right)^2}\right]^{1/2}, \quad (Eq.5)$$

with $Q_1 = \Omega_1/\gamma_1$ is the quality factor of control mode. It is evident that the mechanical gain $G_1 = \frac{|x_1|_{on}}{|x_1|_{off}}$ can be reduced up to a factor of $\frac{1}{2}$, and beyond a threshold of parametric modulation strength ($\delta k_{opt,1} = m_1\Omega_1^2/Q_1$), the gain diverges. In the following squeezing and transfer experiments, we add a weak microwave tone at the cavity resonance and modulate its amplitude with a 1 MHz-wide white noise, to increase the noise amplitudes of the mechanical oscillators up to a detectable level. This excitation noise is sufficiently detuned from other pump tones by $\frac{\kappa}{2\sqrt{3}} \gg \Omega_j$ ($j = 1,2$), ensuring that it does not induce unwanted dynamics in the mechanical modes.

We apply degenerate parametric amplification to the control mode to generate squeezing at $\phi_s = 0$ (**Fig. 2**). **Fig. 2a** shows that as the squeezing modulation $\beta_s$ increases, the gain $G_1$ of the squeezed quadrature ($X_1$) and the anti-squeezed quadrature ($Y_1$) follow the expected parametric gains in Eq (5). **Fig. 2b** shows the plot of $G_1$ as a function of the phase $\phi_1$ for different values of $\beta_s$. This result also aligns well with Eq. (5), clearly illustrating that both squeezing and amplification become more pronounced as $\beta_s$ increases. **Fig. 2c** and **2d** present examples of mechanical fluctuations in the quadrature space when $\beta_s$ is 0.0338 and zero, as a demonstration of squeezed noise profiles. **Fig. 2e** presents the histograms of $X_1$ and $Y_1$ quadrature at various squeezing conditions, showing quantitatively that the squeezed and amplified quadrature has 0.57 times smaller and 3.9 times larger variance compared to that when the parametric amplification is inactive.

**Generation and transfer of mechanical noise squeezing**

The concept of reservoir engineering for mechanical noise squeezing has been demonstrated in cavity optomechanics[5–7]. In these bichromatic drive setups, the cavity is prepared with vacuum noise, and two pump tones are applied. The Bogoliubov modes of mechanical oscillator interact with the cavity photons through a beam-splitter interaction in the rotating frame. In the laboratory frame, this interaction leads to phase-dependent reduction of noise in the mechanical mode, driving the mechanical oscillator into a squeezed state[38]. A different approach for generating mechanical squeezing is directly employing a squeezed reservoir in the laboratory frame. In general, this squeezing transfer refers to the process of inducing squeezing in a bosonic mode via interaction with a squeezed reservoir[16–23]. A beam-splitter or transfer interaction between the mechanical mode and the squeezed reservoir results in the mechanical oscillator in a squeezed state.

Here, the squeezed reservoir is one of the mechanical modes (control), which is coupled to the other mode (target). We apply degenerate parametric amplification to the control mode to squeeze it, while simultaneously activating the transfer interaction between the control and target modes, enabling the target mode to become squeezed. A strong microwave drive, detuned from the cavity resonance, is amplitude-modulated at two frequencies according to **Eq. 2**. By varying the squeezing strength $\beta_s$ while holding the transfer strength $\beta_t \approx 0.28$ constant, we map out the squeezing transfer dynamics.

We have implemented squeezing transfer between two mechanical modes in our cavity optomechanical system. In our two-mechanical-mode system, squeezing transfer is analyzed by computing the full covariance matrix of the quadrature operators. We define the canonical vector $\vec{q} = (\hat{X}_1, \hat{X}_2, \hat{Y}_1, \hat{Y}_2)^T$ and assume that the equation of motion takes the form $\dot{\vec{q}} = A\vec{q} + \hat{n}(t)$ where $A$ is drift matrix and $n$ represents the dissipation contribution. The steady-state covariance matrix is obtained as the solution of the Lyapunov equation $\frac{dV}{dt} = AV + VA^T + D = 0$ (see Supplementary information)[23,39]. Since the microwave drive is continuous rather than pulsed, our measurements probe the steady-state response of the system governed by the effective Hamiltonian (**Eq. 3**), including dissipation.

The gains $G_j$ of the target and control mode quadratures are shown in **Fig. 3a** and **3b**. The fitting curves are obtained from theoretical calculations using the covariance matrix of the two mechanical modes $V_1$. The target mode experiences noise squeezing through the transfer interaction to the control mode, even though it does not directly experience the squeezing interaction. For the target mode, the gain of the amplified quadrature generally increases, while the gain of the squeezed quadrature decreases as the squeezing interaction in the control mode becomes stronger. In addition, the transfer interaction reduces the squeezing in the control mode. This effect is particularly pronounced in the parametric instability region of the control mode (gray area in **Fig. 3a**), where the transfer interaction prevents parametric instability and raises the threshold to a higher $\beta_s$. These observations are consistent with the assumption that the squeezing observed in the target mode originates from the control mode.

**Correlation between two mechanical modes in squeezing transfer process**

**Fig. 4** displays the normalized covariance matrix of the two-mechanical-mode system $(X_1, Y_1, X_2, Y_2)$ which represents the quadrature correlations generated in the parametric squeezing process[40,41]. We analyze the correlations between the quadratures of both modes during the squeezing transfer experiments and find that these correlations display clear distinction between the squeezing transfer and the other two-mode squeezing techniques[11,17,40,42]. The correlation between two quadratures $q_i$ and $q_j$, with $\vec{q} = (X_1, Y_1, X_2, Y_2)^T$, is defined by the correlation matrix as $C_{ij} = \frac{V_{ij}}{\sqrt{V_{ii}}\sqrt{V_{jj}}}$, where $V_{ij}$ is the element of covariance matrix $\left(= \frac{1}{2}\langle(\hat{q}_i - \langle\hat{q}_i\rangle)(\hat{q}_j - \langle\hat{q}_j\rangle) + (\hat{q}_j - \langle\hat{q}_j\rangle)(\hat{q}_i - \langle\hat{q}_i\rangle)\rangle\right)$, quantifying the normalized covariance between two quadratures of two-mechanical mode systems.

We fix the strengths of the squeezing and transfer interactions and observe how the squeezing axes of the mechanical quadrature noise rotate as the modulation phases vary. The squeezing axis angle $\theta$ is defined as the angle between the squeezed quadrature axis and the corresponding $X_j$-axis. We carefully maintain a consistent global phase reference across all instruments, enabling precise verification of the phase relationship between the two mechanical modes (See Methods). **Fig. 4a** illustrates the rotation of the squeezing axes with respect to the squeezing modulation phase $\phi_s$ and transfer modulation phase $\phi_t$. The observed behavior agrees with the effective Hamiltonian in Eq. 3, which is expressed as $\widehat{H}_{eff} = \hbar \left[ \frac{\beta_s \delta\Omega_{opt,1}}{4} (\hat{c}_1^\dagger \hat{c}_1^\dagger + \hat{c}_1 \hat{c}_1) + \frac{\beta_t \sqrt{\delta\Omega_{opt,1}\delta\Omega_{opt,2}}}{2} (\hat{c}_1^\dagger \hat{c}_2 + \hat{c}_2^\dagger \hat{c}_1) \right]$ where all phase dependence is absorbed into the rotated operators $\hat{c}_1(\phi_s, \phi_t) = \hat{b}_1 e^{-\frac{i\phi_s}{2}}$, $\hat{c}_2(\phi_s, \phi_t) = \hat{b}_2 e^{-i\left(\frac{\phi_s}{2} - \phi_t\right)}$. We set the phase reference at $\phi'_s = \phi'_t = 0$ so that the squeezing axes of both mechanical modes are aligned with the $X_j$-axis. Accordingly, we define the phases as

$\phi'_s = \frac{\pi}{2} + \phi_s$ and $\phi'_t = \frac{\pi}{2} + \phi_t$. As $\phi'_s$ varies, the squeezing axes of both the control and target modes rotate by half of the variation in $\phi'_s$. This simultaneous rotation reflects the origin of the target-mode squeezing from the control mode. Variations in $\phi'_t$ only affect the target mode: while the control-mode squeezing axis remains fixed, the squeezing axis of the target mode rotates by an angle equal to $\phi'_t$. From the Lyapunov steady-state solution $V$, the reconstructed Gaussian Wigner function shows that the principal squeezing axis rotates with $\phi_t$ and $\phi_s$, in quantitative agreement with the fit in Fig. 4a (see Supplementary Information).

**Fig. 4 b–d** show the correlation matrix as the transfer modulation phase $\phi_t$ is swept, with $\beta_s = 0.0476$, $\beta_t \approx 0.28$ and $\phi'_s = 0$ held constant. **Fig. 4b** plots selected elements of the correlation matrix as functions of $\phi'_t$, illustrating how the quadrature correlations evolve with the modulation phase. For all values of $\phi'_t$, nonzero off-diagonal elements appear in the correlation matrix, indicating the emergence of additional correlations between the two mechanical modes. This provides evidence that squeezing in the target mode is inherited from the control mode. When $\phi'_s = \phi'_t = 0$, only the diagonal and the off-diagonal elements $C_{14}$, $C_{23}$, $C_{32}$, $C_{41}$ are nonzero, yielding the correlation matrix characteristic of conventional two-mode squeezing[40] or entanglement[11,13].

The effective Hamiltonian in the two-mechanical-mode system (**Eq. 3**) introduces two independent, tunable phases, $\phi_t$ and $\phi_s$, which allow direct control of the correlations without changing the coupling rate. In contrast to the conventional two-mode squeezing Hamiltonian $\hat{b}_1^\dagger \hat{b}_2^\dagger e^{i\phi_e} + \hat{b}_1 \hat{b}_2 e^{-i\phi_e}$, the system provides additional phase tunability, serving as an extra control parameter for shaping the covariance matrix. **Fig. 4c-d** present the correlation matrices from experiments and theoretical predictions for $\phi'_t = -\frac{5\pi}{18}, 0, \frac{4\pi}{18}$ with $\phi'_s = 0$. Near $\phi'_t \approx -\frac{\pi}{4}$, $C_{34}$ is positive while $C_{32}$ is negative; around $\phi'_t \approx +\frac{\pi}{4}$, these correlations invert, with $C_{34}$ becoming negative and $C_{32}$ positive. Moreover, $\pi$- shift in $\phi_t$ flips element of the correlation matrix as $C_{ij}(\phi'_t + \pi) = -C_{ij}(\phi'_t)$ even though the quadrature amplitude data of each mechanical mode remain unchanged. The correlation between the two mechanical quadratures can be continuously tuned, allowing a reversible transition from correlated to anti-correlated behavior and vice versa. Although the evolution of $C_{ij}(\phi'_t)$ demonstrates squeezing transfer between the modes, the large initial thermal occupations ($n \gg 1$) imply the observed correlations arise from thermal fluctuations, and the mechanical modes remain only classically correlated.

**Discussion**

We demonstrate squeezing transfer between two mechanical modes in a three-dimensional microwave cavity optomechanical system. In the unresolved sideband regime, the optomechanical couplings between the cavity mode and each of the mechanical modes translates into an effective coupling between the two mechanical modes, namely control and target. Utilizing this controllable effective interaction between control and target mode, we implement a simultaneous noise squeezing of control mode and squeezing transfer from the control to the target mode. Consequently, single-mode squeezing is observed within each mode, while two-mode squeezing develops between the control and target modes. By analyzing the quadrature correlations of both modes, we confirm that the noise squeezing transfers from control to target, and it also produces highly correlated mechanical quadratures with combined features of both conventional single-mode and two-mode squeezing.

Squeezing is considered as an important resource in precision sensing[43–50], and its distribution offers significant potentials as it could offer a simultaneous noise reduction in multiple mechanical oscillators employed in a force sensing task. In addition, the reentrant cavity structure in this work has been considered for precision measurements of Casimir forces[24], gravitational waves[51], and dark matter[52,53], suggesting possible quantum sensing applications based on this squeezing transfer technique.

Our approach could be extended to a network of more than two mechanical modes to generate $N(>2)$ number of squeezed mechanical modes exhibit mutual correlations. This capability enables distributed quantum sensing (DQS)[54], in which a single squeezed mechanical mode is distributed among $N$ mechanical modes; in our case, distributing phonons in this way should yield Heisenberg scaling, with the measurement sensitivity improving proportionally to $1/N$[55]. Moreover, a phase- and correlation-controlled mechanical array of multiple squeezers offers a new platform for phonon-based quantum information processing, including measurement-based quantum computing[56] and quantum teleportation[57] and could serve as a versatile mechanical simulator[58].

**Methods**

**Device design and fabrication**

The reentrant cavity optomechanical system consists of the bottom and top sections of the cavity, with a membrane positioned at the center of the top section. The cavity is fabricated from aluminum 6061 using conventional machining techniques. Given machining tolerances of approximately ±50 μm, the initial gap of the reentrant cavity is designed to be 100 μm. After machining, the cavity undergoes polishing with fine sandpaper and a polishing compound to minimize cavity dissipation. To reduce the vacuum gap between the tip of the reentrant cavity and the membrane, a fixed linear actuator pushes the bottom of the cavity. Simulations using COMSOL MULTIPHYSICS indicate that the bottom of the cavity must be thinner than 100 μm to accommodate the force applied by the linear actuator. To achieve this, the processed cavity bottom undergoes additional machining, and a specific region is submerged in aluminum etchant at 50 °C for approximately 5 hours, reducing its thickness below 100 μm.

**Measurement setup**

The device is housed at room temperature in a vacuum chamber maintained below $10^{-4}$ Pa. A very weak microwave tone at the cavity resonance, amplitude-modulated with 50 %-depth white noise over a 1 MHz bandwidth, provides uniform low-level pumping of all mechanical modes. A second, strong microwave tone—detuned from the cavity by $\Delta = -\frac{\kappa}{2\sqrt{3}}$—is simultaneously modulated at the transfer and squeezing interaction frequencies, thereby producing a time-varying radiation-pressure drive. The strong tone is amplified by a power amplifier before entering the cavity; RF isolators are installed in front of every source to suppress back-reflections and prevent unintended signal mixing. Output signals are analyzed either with two phase-locked vector network analyzers (operated in continuous-wave mode) or with a lock-in amplifier capable of recording two channels simultaneously.

**Data availability**

The data that support the findings of this study are available from the corresponding author upon reasonable request.

**Code availability**

The custom code used for data analysis and numerical simulations in this study is available from the corresponding author upon reasonable request. Third-party package lists and version information will be provided to support reproducibility.

# Figures

**Fig. 1.** Microwave cavity optomechanical system

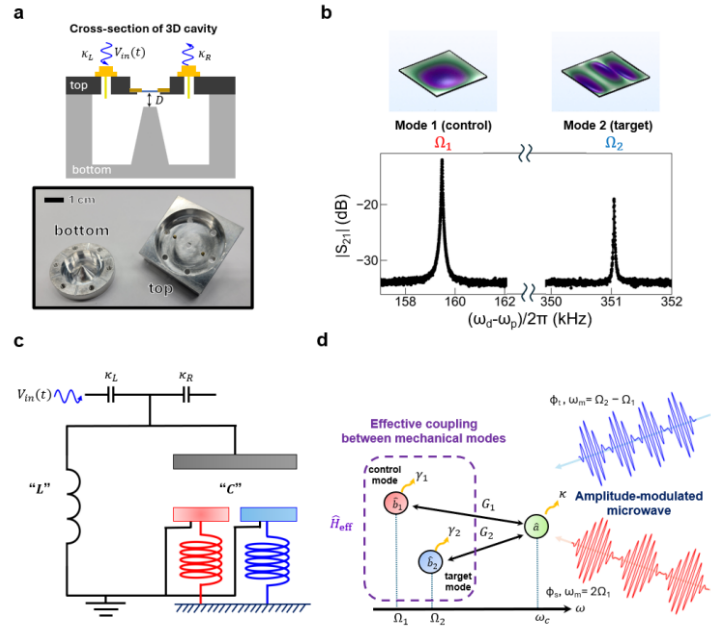

**a.** Device schematic and photographs. The device is composed of a three-dimensional reentrant microwave cavity with a capacitance modulated by the membrane motion. The silicon nitride membrane is metallized with aluminum and positioned above the tip of the reentrant cavity by the vacuum gap (D) to provide the necessary optomechanical coupling between the cavity and mechanical modes. The vacuum gap is adjusted by an external nanoactuator (not shown) up to a few micrometers.

**b.** Optomechanically induced transparency (OMIT) from mechanical resonances. Among the mechanical resonances supported by the membrane, two modes with the largest optomechanical coupling are labeled as mode 1 (control) and mode 2 (target). These modes correspond to the fundamental and (1,3) mode respectively as depicted in the insets.

**c.** Equivalent circuit of **a**. The displacements of the two mechanical oscillators modulate the capacitance, thereby shifting the resonance frequency of the LC resonator.

**d.** Schematic of the cavity optomechanical system comprising two mechanical modes and a single cavity mode. To implement squeezing transfer, the optomechanical system is driven by two microwave tones: one modulated at $2\Omega_1$ and the other at $\Omega_2 - \Omega_1$.

**Fig.2.** Noise squeezing of control mode

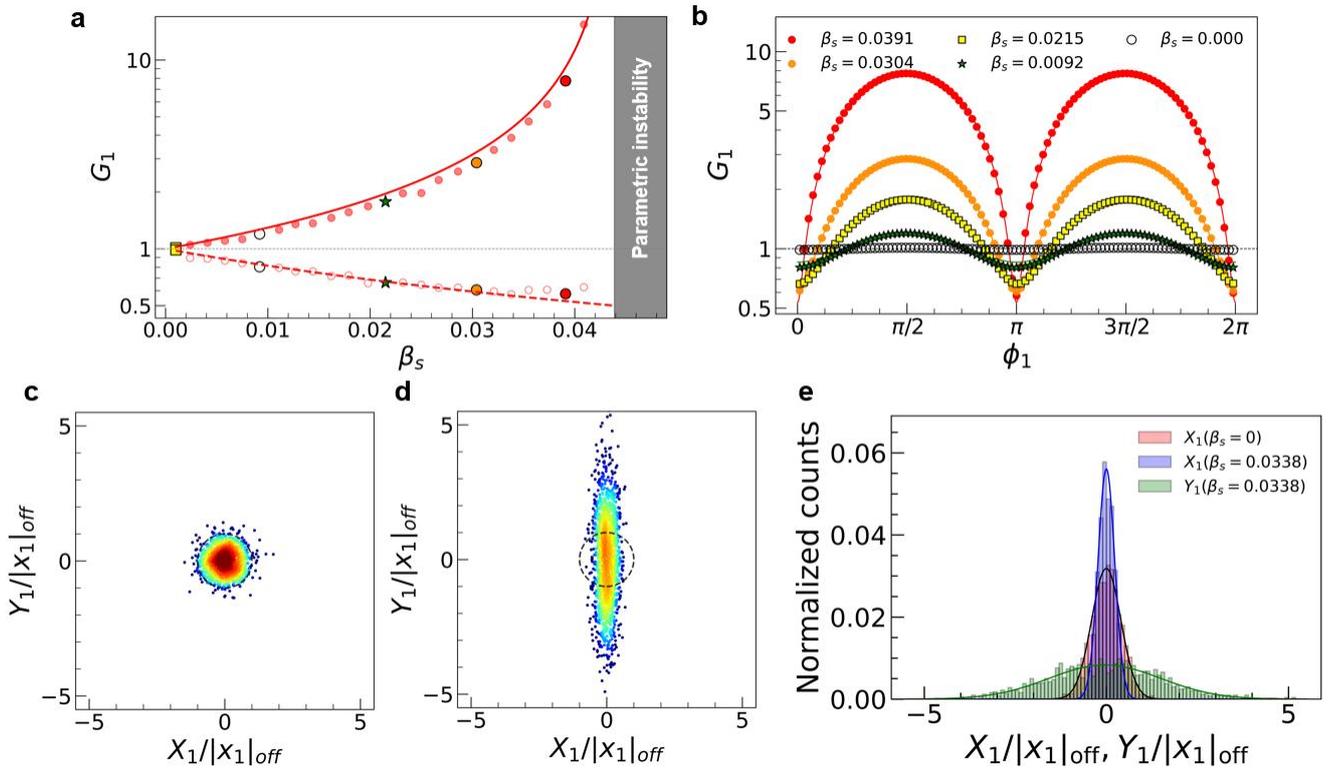

**a**. Gain of squeezed and anti-squeezed quadratures of control mode ($G_1$) vs. $\beta_s$

**b**. $G_1$ vs. $\phi_1$ at various $\beta_s$. The points represent the measured data, while the solid lines correspond to the fitted curves.

**c**. Mechanical quadrature noise when the squeezing modulation is turned off.

**d**. Mechanical quadrature noise when the squeezing modulation is turned on ($\beta_s = 0.0338$). The dashed circle corresponds to $X_1^2 + Y_1^2 = |x_1|_{\text{off}}^2$ serving as a guide to compare **c** and **d**.

**e.** Histogram of quadrature amplitudes when squeezing modulation is turned on ($\beta_s = 0.0338$) and off.

**Fig.3.** Simultaneous squeezing in control and target mode

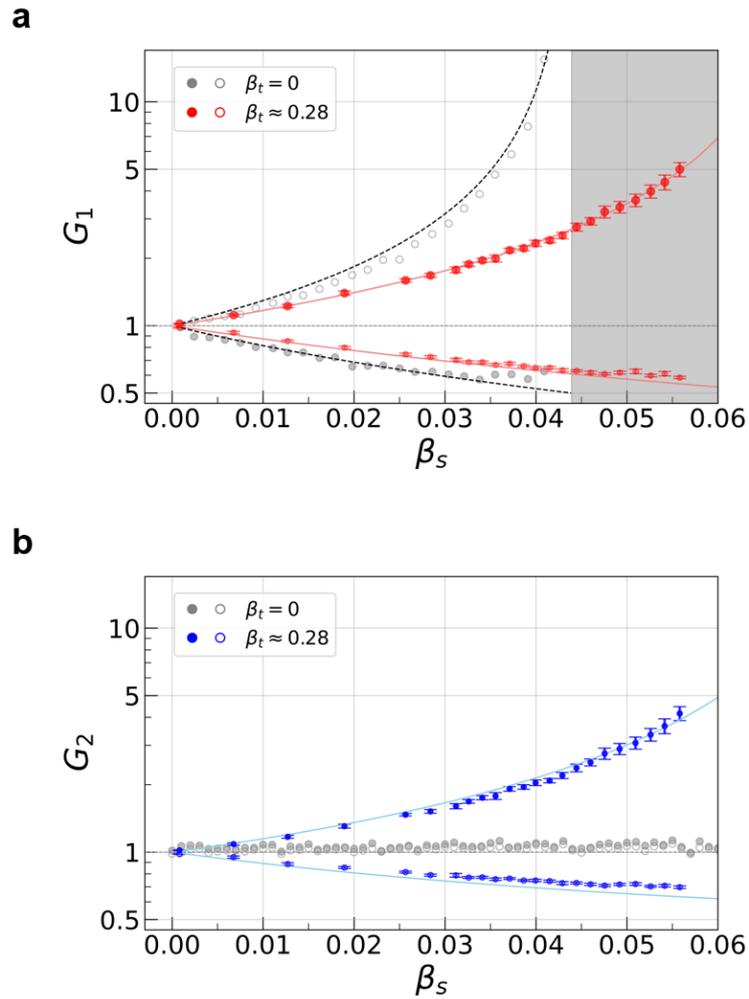

**a.** Gain of squeezed and anti-squeezed quadratures of control mode $(G_1)$ vs. squeezing modulation $(\beta_s)$. The red solid and open circles represent data with the transfer modulation active ($\beta_t \approx 0.28$), while the gray solid and open circles correspond to data with the transfer modulation inactive ($\beta_t = 0$). The gray area indicates the parametric instability regime when the transfer modulation is off.

**b.** Gain of squeezed and anti-squeezed quadratures of target mode $(G_2)$ vs. squeezing modulation depth $(\beta_s)$. The blue solid and open circles represent data with the transfer modulation active ($\beta_t \approx 0.28$), while the gray solid and open circles correspond to data with the transfer modulation inactive ($\beta_t = 0$). The error bars of blue solid and open circles represent the 90% confidence intervals calculated from 12 repeated measurements of the same experiment.

**Fig.4.** Correlation between quadratures of two mechanical modes

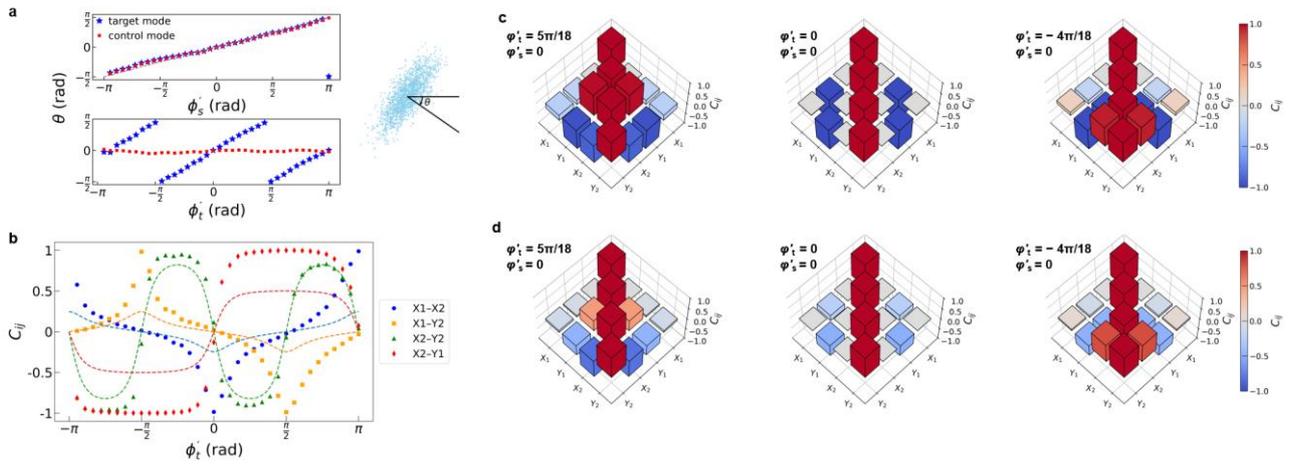

**a.** (Top) squeezing axis angle vs. squeezing modulation phase. The squeezing axes of both the control and target modes rotate by $\varphi_s/2$. (Bottom) squeezing axis angle vs. transfer modulation phase. The squeezing axis of the target mode rotates by $\varphi_t$ whereas the squeezing axis of the control mode remains constant regardless of $\varphi_t$.

**b.** $\phi_t$ vs. correlation coefficients $C_{ij}$. Data points show the measured correlation coefficients for quadrature pairs $X_1 - X_2$, $X_1 - Y_2$, $X_2 - Y_2$ and $X_2 - Y_1$ (blue, yellow, green and red, respectively). Dotted lines are the theoretical predictions.

**c, d.** Correlation matrices $C_{ij}$ of the two-mechanical-mode system are shown for transfer modulation phases $\phi_t = -\frac{5\pi}{18}, 0, \frac{4\pi}{18}$ at fixed $\beta_s = 0.0476$, $\beta_t = 0.28$ and $\phi_s = 0$. **c.** $C_{ij}$ reconstructed from measured data. **d.** Theoretical predictions obtained from the steady-state covariance matrix.

# References

1. Hertzberg, J. B. *et al.* Back-action-evading measurements of nanomechanical motion. *Nature Phys* **6**, 213–217 (2010).

2. Suh, J. *et al.* Mechanically detecting and avoiding the quantum fluctuations of a microwave field. *Science* **344**, 1262–1265 (2014).

3. Szorkovszky, A., Brawley, G. A., Doherty, A. C. & Bowen, W. P. Strong Thermomechanical Squeezing via Weak Measurement. *Phys. Rev. Lett.* **110**, 184301 (2013).

4. Pontin, A. *et al.* Squeezing a Thermal Mechanical Oscillator by Stabilized Parametric Effect on the Optical Spring. *Phys. Rev. Lett.* **112**, 023601 (2014).

5. Wollman, E. E. *et al.* Quantum squeezing of motion in a mechanical resonator. (2015).

6. Pirkkalainen, J.-M., Damskägg, E., Brandt, M., Massel, F. & Sillanpää, M. A. Squeezing of Quantum Noise of Motion in a Micromechanical Resonator. *Phys. Rev. Lett.* **115**, 243601 (2015).

7. Lei, C. U. *et al.* Quantum Nondemolition Measurement of a Quantum Squeezed State Beyond the 3 dB Limit. *Phys. Rev. Lett.* **117**, 100801 (2016).

8. Bothner, D. *et al.* Cavity electromechanics with parametric mechanical driving. *Nat Commun* **11**, 1589 (2020).

9. Pontin, A. *et al.* Dynamical Two-Mode Squeezing of Thermal Fluctuations in a Cavity Optomechanical System. *Phys. Rev. Lett.* **116**, 103601 (2016).

10. Ockeloen-Korppi, C. F. *et al.* Stabilized entanglement of massive mechanical oscillators. *Nature* **556**, 478–482 (2018).

11. Kotler, S. *et al.* Direct observation of deterministic macroscopic entanglement. *Science* **372**, 622–625 (2021).

12. Palomaki, T. A., Harlow, J. W., Teufel, J. D., Simmonds, R. W. & Lehnert, K. W. Coherent state transfer between itinerant microwave fields and a mechanical oscillator. *Nature* **495**, 210–214 (2013).

13. Palomaki, T. A., Teufel, J. D., Simmonds, R. W. & Lehnert, K. W. Entangling Mechanical Motion with Microwave Fields. *Science* **342**, 710–713 (2013).

14. Cirac, J. I., Zoller, P., Kimble, H. J. & Mabuchi, H. Quantum State Transfer and Entanglement

Distribution among Distant Nodes in a Quantum Network. *Phys. Rev. Lett.* **78**, 3221–3224 (1997).

15. Christandl, M., Datta, N., Ekert, A. & Landahl, A. J. Perfect State Transfer in Quantum Spin Networks. *Phys. Rev. Lett.* **92**, 187902 (2004).

16. Jähne, K. *et al.* Cavity-assisted squeezing of a mechanical oscillator. *Phys. Rev. A* **79**, 063819 (2009).

17. Huang, S. & Agarwal, G. S. Entangling nanomechanical oscillators in a ring cavity by feeding squeezed light. *New J. Phys.* **11**, 103044 (2009).

18. Hammerer, K. *et al.* Strong Coupling of a Mechanical Oscillator and a Single Atom. *Phys. Rev. Lett.* **103**, 063005 (2009).

19. Wallquist, M. *et al.* Single-atom cavity QED and optomicromechanics. *Phys. Rev. A* **81**, 023816 (2010).

20. Li, J., Zhu, S.-Y. & Agarwal, G. S. Squeezed states of magnons and phonons in cavity magnomechanics. *Phys. Rev. A* **99**, 021801 (2019).

21. Molinares, H., Eremeev, V. & Orszag, M. Steady-state squeezing transfer in hybrid optomechanics. Preprint at https://doi.org/10.48550/arXiv.2104.11796 (2021).

22. Zhang, W. *et al.* Generation and transfer of squeezed states in a cavity magnomechanical system by two-tone microwave fields. *Opt. Express* **29**, 11773 (2021).

23. Fan, Z.-Y., Zhu, H.-B., Li, H.-T. & Li, J. Magnon squeezing via reservoir-engineered optomagnomechanics. *APL Photonics* **9**, 100804 (2024).

24. Pate, J. M., Goryachev, M., Chiao, R. Y., Sharping, J. E. & Tobar, M. E. Casimir spring and dilution in macroscopic cavity optomechanics. *Nat. Phys.* **16**, 1117–1122 (2020).

25. Weis, S. *et al.* Optomechanically Induced Transparency. *Science* **330**, 1520–1523 (2010).

26. Teufel, J. D. *et al.* Circuit cavity electromechanics in the strong-coupling regime. *Nature* **471**, 204–208 (2011).

27. Le Floch, J.-M. *et al.* Rigorous analysis of highly tunable cylindrical transverse magnetic mode re-entrant cavities. *Review of Scientific Instruments* **84**, 125114 (2013).

28. C. Carvalho, N., Fan, Y. & Tobar, M. E. Piezoelectric tunable microwave superconducting cavity. *Review of Scientific Instruments* **87**, 094702 (2016).

29. Kumar, S., Kenworthy, M., Ginn, H. & Rojas, X. Optomechanically induced transparency/absorption in a 3D microwave cavity architecture at ambient temperature. *AIP Advances* **14**, 035107 (2024).

30. Aspelmeyer, M., Kippenberg, T. J. & Marquardt, F. Cavity optomechanics. *Rev. Mod. Phys.* **86**, 1391–1452 (2014).

31. Yuan, M., Singh, V., Blanter, Y. M. & Steele, G. A. Large cooperativity and microkelvin cooling with a three-dimensional optomechanical cavity. *Nat Commun* **6**, 8491 (2015).

32. Peterson, G. A. *et al.* Ultrastrong Parametric Coupling between a Superconducting Cavity and a Mechanical Resonator. *Phys. Rev. Lett.* **123**, 247701 (2019).

33. Le, A. T., Brieussel, A. & Weig, E. M. Room temperature cavity electromechanics in the sideband-resolved regime. *Journal of Applied Physics* **130**, 014301 (2021).

34. del Pino, J., Slim, J. J. & Verhagen, E. Non-Hermitian chiral phononics through optomechanically induced squeezing. *Nature* **606**, 82–87 (2022).

35. Mathew, J. P., Pino, J. D. & Verhagen, E. Synthetic gauge fields for phonon transport in a nano-optomechanical system. *Nat. Nanotechnol.* **15**, 198–202 (2020).

36. Rugar, D. & Grütter, P. Mechanical parametric amplification and thermomechanical noise squeezing. *Phys. Rev. Lett.* **67**, 699–702 (1991).

37. Slim, J. J. *et al.* Optomechanical realization of the bosonic Kitaev chain. *Nature* **627**, 767–771 (2024).

38. Kronwald, A., Marquardt, F. & Clerk, A. A. Arbitrarily large steady-state bosonic squeezing via dissipation. *Phys. Rev. A* **88**, 063833 (2013).

39. Vitali, D. *et al.* Optomechanical Entanglement between a Movable Mirror and a Cavity Field. *Phys. Rev. Lett.* **98**, 030405 (2007).

40. Mahboob, I., Okamoto, H., Onomitsu, K. & Yamaguchi, H. Two-Mode Thermal-Noise Squeezing in an Electromechanical Resonator. *Phys. Rev. Lett.* **113**, 167203 (2014).

41. Bergeal, N., Schackert, F., Frunzio, L. & Devoret, M. H. Two-Mode Correlation of Microwave Quantum Noise Generated by Parametric Down-Conversion. *Phys. Rev. Lett.* **108**, 123902 (2012).

42. Ou, Z. Y., Pereira, S. F., Kimble, H. J. & Peng, K. C. Realization of the Einstein-Podolsky-Rosen paradox for continuous variables. *Phys. Rev. Lett.* **68**, 3663–3666 (1992).

43. Stoler, D. Equivalence Classes of Minimum Uncertainty Packets. *Phys. Rev. D* **1**, 3217–3219 (1970).

44. Hollenhorst, J. N. Quantum limits on resonant-mass gravitational-radiation detectors. *Phys. Rev. D* **19**, 1669–1679 (1979).

45. Caves, C. M., Thorne, K. S., Drever, R. W. P., Sandberg, V. D. & Zimmermann, M. On the measurement of a weak classical force coupled to a quantum-mechanical oscillator. I. Issues of principle. *Rev. Mod. Phys.* **52**, 341–392 (1980).

46. Caves, C. M. Quantum-Mechanical Radiation-Pressure Fluctuations in an Interferometer. *Phys. Rev. Lett.* **45**, 75–79 (1980).

47. Braginsky, V. B., Vorontsov, Y. I. & Thorne, K. S. Quantum Nondemolition Measurements. *Science* **209**, 547–557 (1980).

48. Aasi, J. *et al.* Enhanced sensitivity of the LIGO gravitational wave detector by using squeezed states of light. *Nature Photon* **7**, 613–619 (2013).

49. Malnou, M. *et al.* Squeezed Vacuum Used to Accelerate the Search for a Weak Classical Signal. *Phys. Rev. X* **9**, 021023 (2019).

50. Backes, K. M. *et al.* A quantum enhanced search for dark matter axions. *Nature* **590**, 238–242 (2021).

51. Blair, D. *et al.* High Sensitivity Gravitational Wave Antenna with Parametric Transducer Readout. *Phys. Rev. Lett.* **74**, 1908–1911 (1995).

52. McAllister, B. T., Parker, S. R. & Tobar, M. E. 3D lumped LC resonators as low mass axion haloscopes. *Phys. Rev. D* **94**, 042001 (2016).

53. Chakrabarty, S. *et al.* Low frequency, 100–600 MHz, searches with axion cavity haloscopes. *Phys. Rev. D* **109**, 042004 (2024).

54. Guo, X. *et al.* Distributed quantum sensing in a continuous-variable entangled network. *Nat. Phys.* **16**, 281–284 (2020).

55. Giovannetti, V., Lloyd, S. & Maccone, L. Quantum Metrology. *Phys. Rev. Lett.* **96**, 010401 (2006).

56. Roh, C., Gwak, G., Yoon, Y.-D. & Ra, Y.-S. Generation of three-dimensional cluster entangled state. *Nat. Photon.* **19**, 526–532 (2025).

57. Takeda, S., Mizuta, T., Fuwa, M., Van Loock, P. & Furusawa, A. Deterministic quantum teleportation of photonic quantum bits by a hybrid technique. *Nature* **500**, 315–318 (2013).

58. Mahboob, I., Okamoto, H. & Yamaguchi, H. An electromechanical Ising Hamiltonian. *Sci. Adv.* **2**, e1600236 (2016).

**Acknowledgements**

This work was supported by Korea Research Institute for defense Technology planning and advancement (KRIT), funded by Defense Acquisition Program Administration (DAPA) (KRIT-CT-23-031), and by the National Research Foundation of Korea (NRF) and the Institute of Information & Communications Technology Planning & Evaluation (IITP), funded by the Ministry of Science and ICT (RS-2024-00352688, RS-2024-00402302, RS-2024-00408271, RS-2023-00259676, RS-2022-00164799).

**Author contributions**

M.J, and J.S. conceived the experiments. M.J. fabricated the devices and performed the experiments. M.J, Y.R, and J.S. developed theoretical treatment. M.J., and J.S. analyzed the data and prepared the paper. J.S. supervised the project. All the authors contributed to the discussions and paper preparation.

**Competing interests**

The authors declare no competing interests.

# Supplementary Information for "Simultaneous generation and transfer of mechanical noise squeezing"

Mungyeong Jeong[1], Hyojun Seok[2], Young-Sik Ra[3] & Junho Suh[1*]

[1]*Department of Physics, Pohang University of Science and Technology (POSTECH), Pohang, Korea*

[2]*Department of Physics Education, Seoul National University, Seoul, Korea*

[3] *Department of Physics, Korea Advanced Institute of Science and Technology, Daejeon, Korea*

[*]Corresponding author. Email: junhosuh@postech.ac.kr

## SUPPLEMENTARY NOTE 1. SYSTEM PARAMETERS

| Parameter | Symbol | Value |
|---|---|---|
| Cavity resonance frequency | $\omega_c$ | $2\pi \times 3.5035$ GHz |
| Total cavity dissipation rate | $\kappa$ | $2\pi \times 38.9$ MHz |
| External cavity dissipation rate | $\kappa_{ex}$ | $2\pi \times 787$ kHz |
| Mechanical resonance frequency of control mode (mode 1) | $\Omega_1$ | $2\pi \times 159$ kHz |
| Mechanical resonance frequency of target mode (mode 2) | $\Omega_2$ | $2\pi \times 351$ kHz |
| Mechanical dissipation of control mode (mode 1) | $\gamma_1$ | $2\pi \times 39.9$ Hz |
| Mechanical dissipation of target mode (mode 2) | $\gamma_2$ | $2\pi \times 13.8$ Hz |
| Single photon optomechanical coupling rate of control mode (mode 1) | $g_{01}$ | $2\pi \times 28.2$ mHz |
| Single photon optomechanical coupling rate of target mode (mode 2) | $g_{02}$ | $2\pi \times 6.31$ mHz |

## SUPPLEMENTARY NOTE 2. EXPERIMENTAL METHODS

Supplementary Figure 1 illustrates the experimental setup (circuit). The pump tone's amplitude is modulated at a frequency determined by external function generator (TEKTRONIX AFG3102), causing the driven photon number to vary over time and thus modulate the radiation pressure force. To achieve a high-power microwave drive, the amplitude-modulated microwave signal from the generator (Rohde&Schwarz SMP02) is sent through a power amplifier (Mini-Circuits ZVE-3W-83+) prior to injection into the cavity optomechanical system. An additional RF isolator is installed in front of all microwave sources to prevent unexpected signal mixing caused by microwave backflow into the sources, which can occur when a very strong microwave pump is produced by the power amplifier. The cavity optomechanical system is placed in a vacuum chamber at pressures below $10^{-3}$ mTorr, and all measurements are performed at room temperature. The vacuum chamber is mounted on an optical table to isolate it from external vibrations.

The measurements are conducted using two Vector Network Analyzers (VNAs; Agilent Technologies N5230A, N9020A) and a lock-in amplifier (Zurich Instruments HF2LI). The VNA measurements are performed in Continuous Wave Time mode (CW mode), in which a continuous wave at a single frequency is used. In this mode, the probe tone of the VNA is not injected into the optomechanical system; instead, the microwave signal emitted from the system is measured to analyze squeezing noise. Each VNA measures the optomechanical sidebands of the control and target modes, respectively. All measurement instruments are phase-locked to a common 10 MHz reference. However, in the CW mode of the VNA, a new phase reference is established each time a trigger is initiated, causing the phase relationship between the two VNAs to become arbitrary. This issue does not occur in the lock-in amplifier, allowing the phase relationship between the two modes to be determined. The reference tone injected into the lock-in amplifier is measured using an external function generator (TEKTRONIX AFG3022B). We find the noise performance of the lock-in amplifier is inferior to that of the VNAs, and carry noise measurements for squeezing and anti-squeezing gains using the VNAs: Fig. 3(a,b), which shows the measured angle of the squeezing axis of the mechanical quadrature noise, presents data obtained using a lock-in amplifier, whereas all other data were acquired using VNAs.

A pump tone is injected into the optomechanical system detuned by $\Delta = -\frac{\kappa}{2\sqrt{3}}$. As described in the main text, to generate and transfer squeezing in the control mode, the pump tone is amplitude-modulated at frequencies $\Omega_1$ and $\Omega_2 - \Omega_1$. The modulation at these frequencies corresponds to the squeezing and transfer interactions, respectively. These two modulation depths are proportional to the function amplitude set on the function generator driving the microwave input. To further excite the mechanical oscillators, we inject a microwave tone at the cavity resonance with a 1 MHz-wide white noise amplitude modulation and a modulation depth of 50%. This tone weakly beats the mechanical modes, effectively pumping the mechanical oscillators. This white noise does not introduce any additional optical spring effect because it is injected at the cavity center.

**SUPPLEMENTARY NOTE 3. CHARATERIZATION OF MECHANICAL PROPERTIES**

Since our experiment is conducted in a room-temperature laboratory, we directly measure the microwave power reaching the 3D cavity using a spectrum analyzer to calibrate the intracavity driven photon number $n_d$. With the calibrated $n_d$, we extract the single-photon optomechanical coupling rate and the mechanical linewidth using Optomechanically Induced Transparency (OMIT). When the resolved sideband regime is not assumed, the transmission $S_{21}(\omega)$ with an additional single drive follows the equation below[1]:

$$S_{21}(\omega) = \frac{1 + if_j(\omega)}{-i(\Delta + \omega - \omega_d) + \frac{\kappa}{2} + 2\Delta f_j(\omega)} \frac{\kappa_{ex}}{2} \qquad (3.1)$$

where,

$$f_j(\omega) = \hbar n_d \left(\frac{g_0}{x_{zp,j}}\right)^2 \frac{\chi_j(\omega)}{i(\Delta - \omega - \omega_d) + \frac{\kappa}{2}} \qquad (3.2)$$

and

$$\chi_j(\omega) = \frac{1}{m_{eff,j}} \times \left(\frac{1}{\Omega_j^2 - (\omega - \omega_d)^2 - i(\omega - \omega_d)\gamma_{m,j}}\right) \qquad (3.3)$$

When the drive tone is applied at the cavity center, the measured S21 is shown in **Supplementary Figure 3**, and we select the two modes. The (3,1) mode loses its degeneracy with the (1,3) mode due to the asymmetry in the side lengths of the silicon nitride membrane.

## SUPPLEMENTARY NOTE 4. DATA ANALYSIS

The analysis of mechanical quadrature noise begins by rotating the noise data, plotted in the $X_j - Y_j$ quadrature plane, by an angle $\phi$. The rotated data is then projected onto the $X_j$-axis, and a histogram of the projected displacements is constructed. Finally, the histogram is fitted with a Gaussian function to extract the quadrature noise characteristics[2]. In all experiments presented in this paper, the standard deviation of all Gaussian fittings is at least two orders of magnitude smaller than the fitted value. In analyzing the two mechanical-mode quadrature noise simultaneous, we mitigate uncontrolled measurement drift by first rotating the control-mode noise in its X-Y histogram so that its squeezing axis $\theta$ is set to zero, then applying the same rotation to the target-mode noise. Each dataset consists of 1700 points recorded by the lock-in amplifier at 15 ms intervals and 1601 points recorded by the VNA at 91 ms intervals.

## SUPPLEMENTARY NOTE 5. EFFECTIVE MECHANICAL HAMILTONIAN AND SQUEEZING AXIS

The Hamiltonian of our optomechanical system, which consists of two mechanical resonators and a single cavity mode, is given by

$$H = -\hbar\Delta \hat{a}^+\hat{a} + \hbar\sum_{j=1,2}\omega_j \hat{b}_j^+\hat{b}_j - g_{0j}\hat{a}^+\hat{a}(\hat{b}_j^+ + \hat{b}_j) \tag{5.1}$$

Under the assumptions of large detuning ($\Delta \gg \omega_j$) and the fast-cavity limit ($\kappa \gg \omega_j$), we linearize the Hamiltonian ($\hat{a} \to \hat{a} + \delta\hat{a}$ and neglecting terms $\mathcal{O}((\delta\hat{a})^2)$) and apply a second-order perturbation that includes the adiabatic elimination of cavity fluctuation $\delta\hat{a}$[3]. Transforming into the rotating frame at the mechanical frequencies and applying the rotating wave approximation, the resulting effective mechanical Hamiltonian is given by Eq. (3) in the main text[4].

By leveraging the fact that squeezing occurs in the effective Hamiltonian $\hat{H}_{eff}$ (Eq.3), we can determine the rotation angle of the squeezing axes of mechanical modes as a function of $\phi_t$ and $\phi_s$ without explicitly solving the equations of motion. The rotation of the squeezing axis of the mechanical quadrature noise as the modulation phase varies is confirmed by the effective Hamiltonian (**Eq. 3**). Here, we define the rotated operators by introducing an additional phase factor for mode 1 and mode 2 as follows:

$$\hat{c}_1(\phi_s, \phi_t) = \hat{b}_1 e^{-\frac{i\phi_s}{2}} \tag{5.2}$$

$$\hat{c}_2(\phi_s, \phi_t) = \hat{b}_2 e^{-i\left(\frac{\phi_s}{2} - \phi_t\right)} \tag{5.3}$$

In this case, the effective Hamiltonian (**Eq. 3**) is rewritten as:

$$\hat{H}_{eff} = \hbar\left[g_s\left(\hat{b}_1^+\hat{b}_1^+ e^{i\phi_s} + \hat{b}_1\hat{b}_1 e^{-\phi_s}\right) + g_t\left(\hat{b}_1^+\hat{b}_2 e^{i\phi_t} + \hat{b}_2^+\hat{b}_1 e^{-i\phi_t}\right)\right]$$

$$= \hbar[g_s(\hat{c}_1^+\hat{c}_1^+ + \hat{c}_1\hat{c}_1) + g_t(\hat{c}_1^+\hat{c}_2 + \hat{c}_1\hat{c}_2^+) \tag{5.4}$$

Therefore, if we define the squeezing axes of both modes as the $X_j$-axis ($X_j = \frac{1}{\sqrt{2}}(\hat{b}_1 + \hat{b}_1^+)$) when $\phi_s = \phi_t = 0$, then for fixed $\phi_s, \phi_s$, the squeezing axis of the target mode rotates by $\phi_t + \phi_s/2$, while squeezing axis of the control mode rotates by $\phi_s/2$ from the $X_j$-axis.

**SUPPLEMENTARY NOTE 6. STEADY-STATE COVARIANCE MATRIX**

For the canonical vector $\vec{q} = (\hat{X}_1, \hat{X}_2, \hat{Y}_1, \hat{Y}_2)^T$, the case with dissipation can be expressed in the form $\dot{\vec{q}} = A\vec{q} + \hat{n}(t)$ where $A$ is drift matrix and $\hat{n}(t)$ represents the dissipation contribution. By definition, the covariance matrix is given by $V_{ij} = \frac{1}{2}\langle(\hat{q}_i - \langle\hat{q}_i\rangle)(\hat{q}_j - \langle\hat{q}_j\rangle) + (\hat{q}_j - \langle\hat{q}_j\rangle)(\hat{q}_i - \langle\hat{q}_i\rangle)\rangle$, its time derivative can be written as follows.

$$\frac{dV}{dt} = AV + VA^T + D \tag{6.1}$$

According to the input–output relation, the noise term can be written as $\hat{n}(t) = (\sqrt{\gamma_1}X_{1,in}, \sqrt{\gamma_2}X_{2,in}, \sqrt{\gamma_1}Y_{1,in}, \sqrt{\gamma_2}Y_{2,in})^T$, and $D = diag\left(\gamma_1\left(n_1 + \frac{1}{2}\right), \gamma_2\left(n_2 + \frac{1}{2}\right), \gamma_1\left(n_1 + \frac{1}{2}\right), \gamma_2\left(n_2 + \frac{1}{2}\right)\right)$ [5]. Furthermore, from the Hamiltonian ($\hat{H}_{eff} = \hbar\left[g_t(\hat{b}_1^+\hat{b}_2 e^{+i\phi_t} + \hat{b}_2^+\hat{b}_1 e^{-i\phi_t}) + \frac{g_s}{2}(\hat{b}_1^+\hat{b}_1^+ e^{i\phi_s} + \hat{b}_2^+\hat{b}_1 e^{-i\phi_s})\right]$, Eq. 3 in the main text), the corresponding equations of motion are obtained as follows:

$$\dot{\hat{b}}_1 = -\frac{\gamma_1}{2}\hat{b}_1 - ig_t e^{-i\phi_t}\hat{b}_2 + ig_s e^{i\phi_s}\hat{b}_1^+ + \sqrt{\gamma_1}\hat{b}_{1,in} \qquad (6.2)$$

$$\dot{\hat{b}}_2 = -\frac{\gamma_2}{2}\hat{b}_1 - ig_t e^{i\phi_t}\hat{b}_1 + \sqrt{\gamma_2}\hat{b}_{2,in} \qquad (6.3)$$

Therefore, $A$ is obtained as follows:

$$A = \begin{pmatrix} -\frac{\gamma_1}{2} - 2g_s \sin\phi_s & -g_t \sin\phi_t & 2g_s \cos\phi_t & g_t \cos\phi_t \\ g_t \sin\phi_t & -\frac{\gamma_2}{2} & g_t \cos\phi_t & 0 \\ 2g_s \cos\phi_t & -g_t \cos\phi'_t & -\frac{\gamma_1}{2} + 2g_s \sin\phi_s & g_t \sin\phi_t \\ -g_t \cos\phi_t & 0 & -g_t \sin\phi_t & -\frac{\gamma_2}{2} \end{pmatrix} \qquad (6.4)$$

The steady-state $V$ can be directly achieved by solving the Lyapunov equation $\frac{dV}{dt} = AV + VA^T + D = 0$[5,6]. We solve $AV + VA^T + D = 0$ using *Scipy*'s *solve_continuous_lyapunov*, which implements the Bartels–Stewart Schur–based method[7,8]. **Supplementary Figure 4** shows the Wigner function calculated from the obtained covariance matrix $V$[9]. Each plot is normalized by $|x_j|$ with squeezing turned off, as in the main text, along the $X_j$- and $Y_j$- axes. **Supplementary Figure 4 (a)** corresponds to the case where mode 1 is driven by a squeezing interaction stronger than the parametric threshold, so that a steady-state covariance matrix of mode 1 cannot be obtained. **Supplementary Figure 4 (b)** illustrates the case with only the transfer interaction and no squeezing, where both modes are observed to remain unsqueezed. In **Supplementary Figure 4 (c)**, while the squeezing strength of mode 1 on its own would exceed the parametric threshold and induce instability, the presence of the transfer interaction between the two modes suppresses this effect.

To register the instrument-defined quadratures with our theoretical convention, we reparameterize the phases as $\phi'_s \equiv \frac{\pi}{2} + \phi_s$ and $\phi'_t \equiv \frac{3\pi}{2} + \phi_t$. Under this convention, varying $\phi_s'$ and $\phi_t'$ produces a rigid rotation of the Gaussian Wigner ellipse in the $\{X_j, Y_j\}$ plane. In **Supplementary Figure 5**, as $\phi_s'$ and $\phi_t'$ are swept, the Wigner functions computed from the Lyapunov-solution covariance $V$ rotate accordingly, consistent with the expected phase-controlled rotation of the principal squeezing axes.

**SUPPLEMENTARY NOTE 7. PARAMETRIC AMPLIFICATION VIA TIME-MODULATED**

# OPTICAL SPRING EFFECT

Consider applying a drive tone to the optomechanical system that is detuned by $\Delta$ from the cavity center. Specifically, we examine the case where this drive tone is time-modulated at twice the resonant frequency of mechanical mode $j$ $(j = 1,2)$. In this case, the equation of motion for mechanical mode $j$ is given by:

$$m(d^2x)/(dt^2) + m\gamma\, dx/dt + k(t)x = F\cos(\Omega t + \phi) \tag{7.1}$$

In this section, the subscript j is omitted for clarity and to prevent confusion. The spring constant of mechanical mode is given by:

$$k(t) = m\left(\Omega + \delta\Omega_{\text{opt}}(t)\right)^2 \approx m\left(\Omega + g^2 \frac{2\Delta}{\frac{\kappa^2}{4} + \Delta^2}(1 + \beta_s \sin(2\Omega t - \phi_s))\right)^2$$

$$\approx m\left(\Omega^2 + g^2 \frac{4\Omega\Delta}{\frac{\kappa^2}{4} + \Delta^2}(1 + \beta_s \sin(2\Omega t - \phi_s))\right) \tag{7.3}$$

The spring constant of mechanical mode $j$ can be decomposed into a static component $k_0$ and a time-modulating component $k_p$ as follows $k_j(t) = k_0 + k_p \sin(2\Omega t - \phi_s)$ where $k_0 = m\Omega^2 + mg^2 \frac{4\Omega\Delta}{\frac{\kappa^2}{4} + \Delta^2}$ and $k_p = mg^2 \frac{4\Omega\Delta}{\frac{\kappa^2}{4} + \Delta^2} \beta_s$. With the transformations $\widetilde{\Omega} = \Omega\left[\left(1 - \frac{1}{4Q_m^2}\right)^{\frac{1}{2}} + \frac{j}{2Q_m}\right]$, $A = \frac{dx}{dt} + j\widetilde{\Omega}^* x$, $A^* = \frac{dx}{dt} - j\widetilde{\Omega}^* x$.

Then, we rewrite the equation of motion as

$$\dot{A} = j\widetilde{\Omega}A + j\frac{k_p \sin(2\Omega t - \phi_s)}{m_{eff}}\frac{A - A^*}{\widetilde{\Omega}^* + \widetilde{\Omega}} + \frac{F}{m_{eff}}\cos(\Omega t + \phi) \tag{7.4}$$

With the Ansatz $A = A_0 e^{i\Omega t}$ and the high-$Q_m$ $(\Omega \gg \gamma)$ approximations ($\widetilde{\Omega}^* + \widetilde{\Omega} \cong 2\Omega$, $\widetilde{\Omega} - \Omega \cong j\frac{\Omega}{2Q_m}$), we find in rotating wave approximation

$$\frac{\Omega}{2Q_m}A_0 + \frac{k_p}{4m_{eff}\Omega_m}A_0^* - \frac{F_0}{2m_{eff}}e^{i\phi} = 0 \tag{7.5}$$

we can solve this equation for the real and imaginary part of $A_0$. Setting $x(t) = X\cos(\Omega t) + Y\sin(\Omega t)$ and using $X = \text{Im}(A_0)/\Omega_m$, $Y = \text{Re}(A_0)/\Omega_m$

$$X = F_0 \frac{Q_m}{k_0} \left[ \frac{\left(1 + \frac{Q_m k_p}{2k_0}\right) \sin(\phi - \phi_s)}{1 - \frac{Q_m^2 k_p^2}{4k_0^2}} \right] \quad (7.6)$$

$$Y = F_0 \frac{Q_m}{k_0} \left[ \frac{\left(1 - \frac{Q_m k_p}{2k_0}\right) \cos(\phi - \phi_s)}{1 - \frac{Q_m^2 k_p^2}{4k_0^2}} \right] \quad (7.7)$$

From this we can calculate the mechanical amplitude as $|x| = \sqrt{X^2 + Y^2}$ and get

$$|x|_{on} = |x|_{off} \left[ \frac{\cos^2(\phi + \phi_s)}{\left(1 + \frac{Q_m k_p}{2k_0}\right)^2} + \frac{\sin^2(\phi + \phi_s)}{\left(1 - \frac{Q_m k_p}{2k_0}\right)^2} \right]^{1/2} \quad (7.8)$$

Where the amplitude without parametric drive is given by

$$|x|_{off} = F_0 \frac{Q_m}{k_0} \quad (7.9)$$

describing the square root of a Lorentzian line around the resonance frequency.

The squeezing gain of mechanical noise $G$ is

$$G = \left[1 + \frac{Q_m k_p}{2k_0}\right]^{-1} = \left[1 + \frac{\Omega\left(g^2 \frac{4\Omega\Delta}{\frac{\kappa^2}{4} + \Delta^2} \beta_s\right)}{2\gamma_m \left(m\Omega^2 + g^2 \frac{4\Omega\Delta}{\frac{\kappa^2}{4} + \Delta^2}\right)}\right]^{-1} = \left[1 + \frac{\left(\frac{\delta\Omega_{opt}}{\gamma_m}\right)}{1 + 2\left(\frac{\delta\Omega_{opt}}{\Omega}\right)} \beta_s\right]^{-1} = \left(1 + \frac{\beta_s}{\beta_{th}}\right)^{-1} \quad (7.10)$$

$\beta_{th} = \frac{1 + 2(\delta\Omega_{opt}/\Omega_m)}{\left(\frac{\delta\Omega_{opt}}{\gamma_m}\right)}$ is the threshold of squeezing strength.

**SUPPLEMENTARY NOTE 8. NUMERICAL SIMULATION OF SQUEEZING TRANSFER**

It is clear from the effective mechanical Hamiltonian (Eq. 3) that mode 2 undergoes a squeezing interaction through the swapping interaction. Nonetheless, we have additionally confirmed this theoretically using the classical equations of motion. When mode 1 is subjected to degenerate parametric amplification and a transfer interaction is present between mode 1 and mode 2, the classical equations of motion are given by ($m_1 \approx m_2 = 1$):

$$\ddot{x}_1 + \gamma_1 \dot{x}_1 + \Omega_1^2 x_1 + \eta_1 \sin(2\Omega_1 t) x_1 + \Lambda \cos[(\Omega_2 - \Omega_1)t] x_2 = F_1(t) \tag{8.1}$$

$$\ddot{x}_2 + \gamma_2 \dot{x}_2 + \Omega_2^2 x_2 + \Lambda \cos[(\Omega_2 - \Omega_1)t] x_1 = F_2(t) \tag{8.2}$$

Since $x_j(t) = X_j \cos(\Omega_j t) + Y_j \sin(\Omega_j t)$, we substitute this expression into the equations and simplify accordingly. In this process, we apply the rotating wave approximation, assuming that the slowly varying components $X_j(t)$ and $Y_j(t)$ have negligible second derivatives and neglecting all rapidly oscillating terms. After substitution and simplification, we obtain:

$$\ddot{x}_2 + \gamma_2 \dot{x}_2 + \Omega_2^2 x_2 + \Lambda \cos[(\Omega_2 - \Omega_1)t] x_1 \approx \cos(\omega_2 t) \left[ 2\Omega_2 Y_2' + \gamma_2 X_2' + \gamma_2 \Omega_2 Y_2 + \frac{\Lambda}{2} X_1 \right]$$
$$+ \sin(\omega_2 t) \left[ -2\Omega_2 X_2' + \gamma_2 Y_2' - \gamma_2 \Omega_2 X_2 + \frac{\Lambda}{2} Y_1 \right] = F_2(t) \tag{8.3}$$

$$\ddot{x}_1 + \gamma_1 \dot{x}_1 + \Omega_1^2 x_1 + \eta_1 \sin(2\Omega_1 t) x_1 + \Lambda \cos[(\Omega_2 - \Omega_1)t] x_2$$
$$\approx \cos(\omega_1 t) \left[ 2\Omega_1 Y_1' + \gamma_1 X_1' + \gamma_1 \Omega_1 Y_1 + \frac{\eta}{2} Y_1 + \frac{\Lambda}{2} X_2 \right]$$
$$+ \sin(\omega_1 t) \left[ -2\Omega_1 X_1' + \gamma_1 Y_1' - \gamma_1 \Omega_1 X_1 + \frac{\eta}{2} X_1 + \frac{\Lambda}{2} Y_2 \right] = F_1(t) \tag{8.4}$$

Using these equations, the system can be solved numerically. We model the Langevin force as random Gaussian noise, and obtained 1600 data points[2,10] (**Supplementary Figure 6**). The Langevin force, treated as white noise, is assigned different magnitudes for each mode to account for the optomechanical coupling rates. Fig. S4 presents the simulation data obtained by varying $\eta$ while keeping $\Lambda$ fixed. The results indicate that single-mode squeezing occurs simultaneously in both modes at comparable levels.

**Figures**

**Supplementary Figure 1.** Experimental circuit

MWG: microwave generator; FG: function generator; VNA: vector network analyzer; SA: spectrum analyzer; MIX: mixer; LO: local oscillator.

**Supplementary Figure 2.** Microwave setup

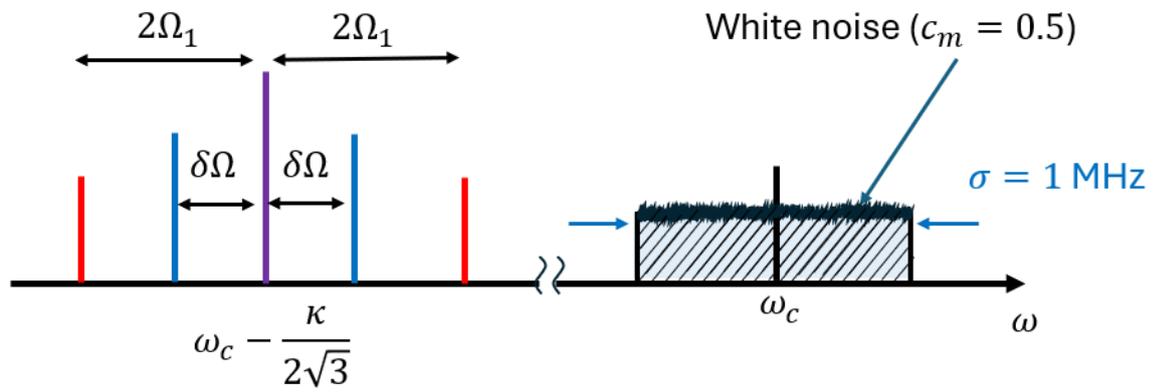

Injected microwave tones: the blue line represents the transfer modulation, and the red line represents the squeezing modulation. Line lengths are not drawn to scale.

**Supplementary Figure 3.** Optomechanically induced transparency

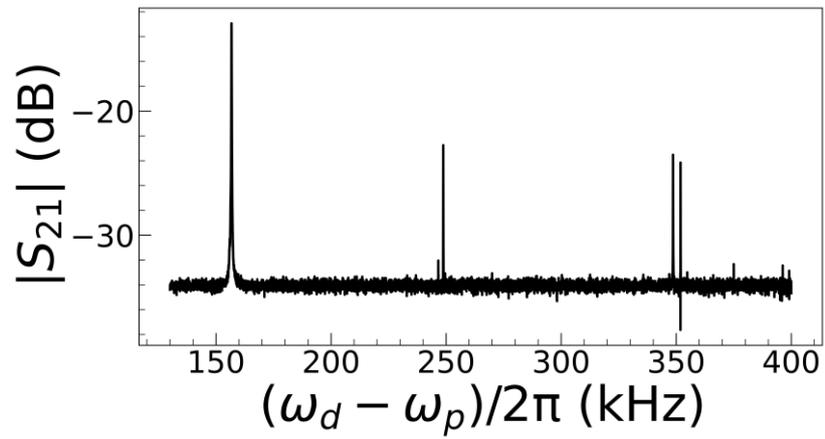

When the drive frequency is set to the cavity resonance, optomechanically induced transparency is observed for all mechanical modes, and no additional white noise is injected. $\omega_d$ denotes the drive frequency, set at the cavity resonance, while $\omega_p$ denotes the probe frequency.

**Supplementary Figure 4.** Wigner function of each mode (left: mode 1, right: mode 2).

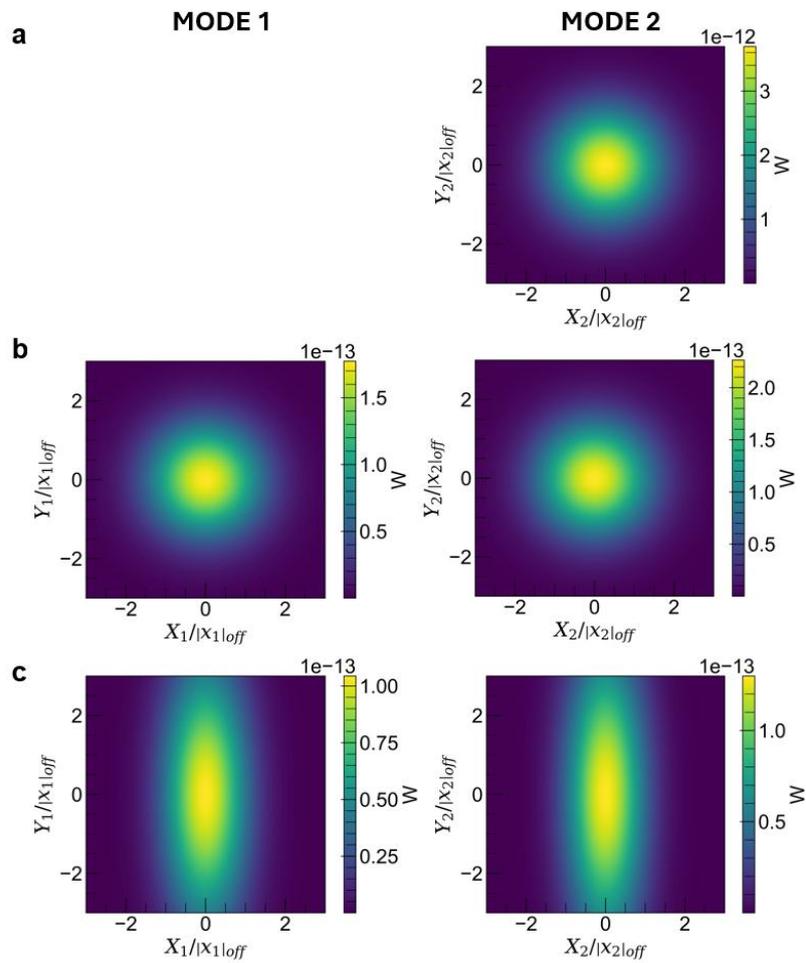

(a) $\beta_s = 0.0476$, $g_t = 0$ (squeezing interaction on, transfer interaction off)

(b) $\beta_s = 0$, $\beta_t = 0.28$ (squeezing interaction off, transfer interaction on)

(c) $\beta_s = 0.0476$, $\beta_t = 0.28$ (squeezing interaction on, transfer interaction on)

**Supplementary Figure 5.** Wigner functions of the two mechanical modes in the squeezing-transfer condition (left, mode 1; right, mode 2)

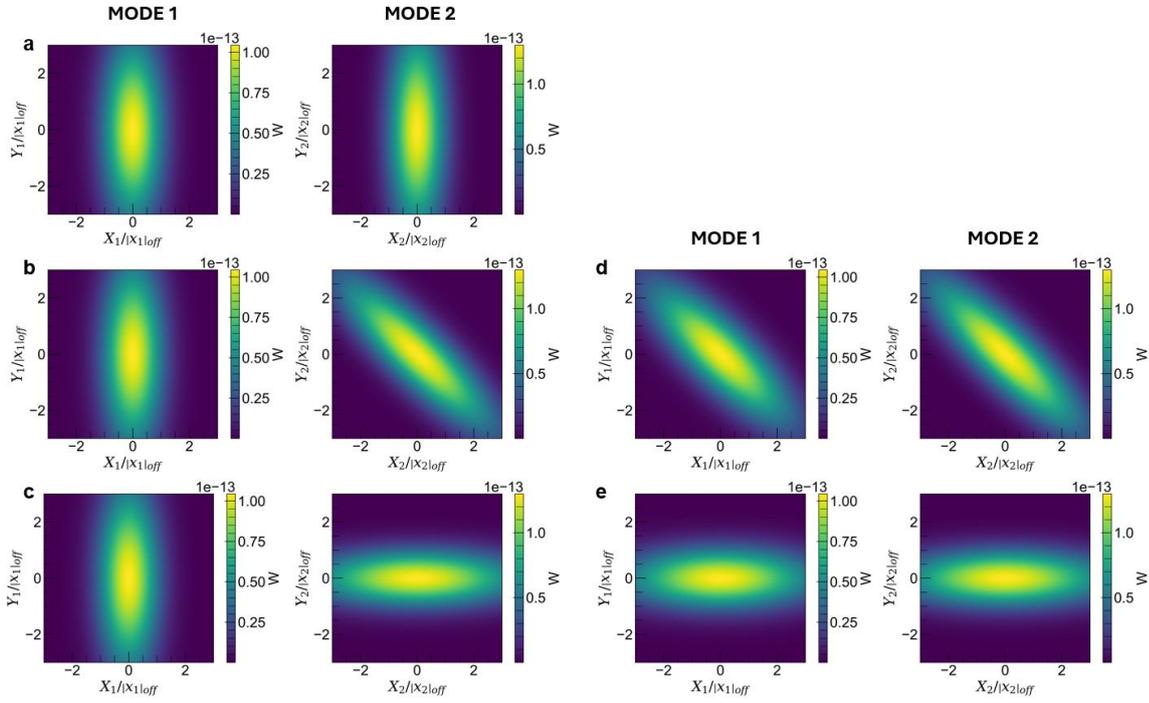

(a) $\phi_s' = 0$, $\phi_t' = 0$

(b) $\phi_s' = 0$, $\phi_t' = \pi/4$

(c) $\phi_s' = 0$, $\phi_t' = \pi/2$

(d) $\phi_s' = \pi/2$, $\phi_t' = 0$

(e) $\phi_s' = \pi$, $\phi_t' = 0$

Both squeezing and transfer interactions are on in all cases.

**Supplementary Figure 6.** Transfer squeezing: Numerical Calculation

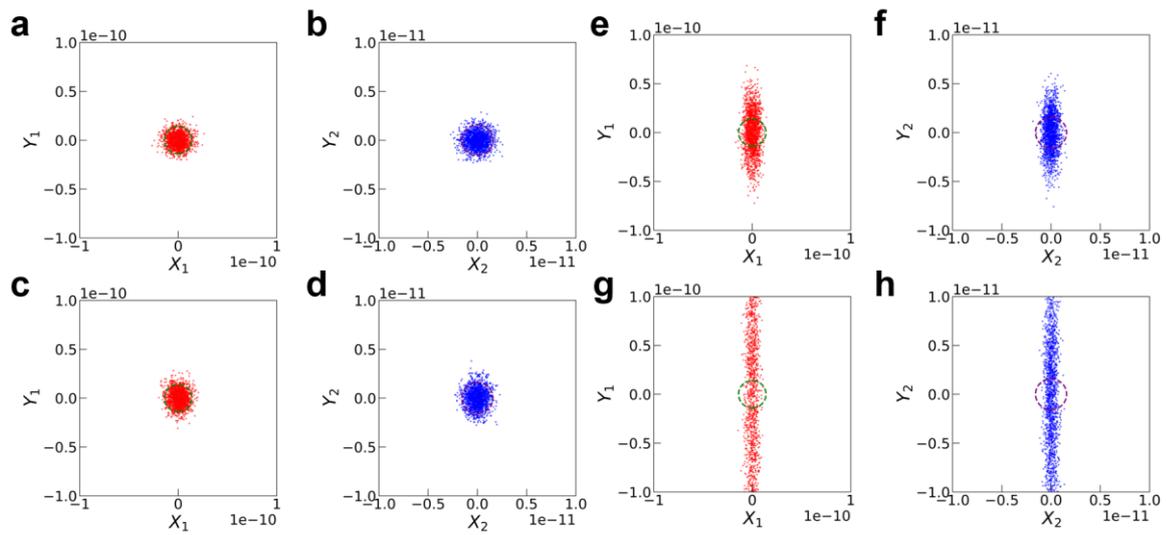

The transfer squeezing is calculated through numerical simulations. The transfer strength $\Lambda$ is fixed at $10^8$, while the squeezing strength $\eta$ is varied as 0 (a, b), $10^8$ (c, d), $5\times10^8$ (e, f), and $9\times10^8$ (g, h). The results show that as the squeezing strength increases, squeezing appears in both the control and target modes. The dashed circles are drawn for comparison.

# References

1. Weis, S. *et al.* Optomechanically Induced Transparency. *Science* **330**, 1520–1523 (2010).

2. Mahboob, I., Okamoto, H., Onomitsu, K. & Yamaguchi, H. Two-Mode Thermal-Noise Squeezing in an Electromechanical Resonator. *Phys. Rev. Lett.* **113**, 167203 (2014).

3. Reiter, F. & Sørensen, A. S. Effective operator formalism for open quantum systems. *Phys. Rev. A* **85**, 032111 (2012).

4. del Pino, J., Slim, J. J. & Verhagen, E. Non-Hermitian chiral phononics through optomechanically induced squeezing. *Nature* **606**, 82–87 (2022).

5. Fan, Z.-Y., Zhu, H.-B., Li, H.-T. & Li, J. Magnon squeezing via reservoir-engineered optomagnomechanics. *APL Photonics* **9**, 100804 (2024).

6. Vitali, D. *et al.* Optomechanical Entanglement between a Movable Mirror and a Cavity Field. *Phys. Rev. Lett.* **98**, 030405 (2007).

7. Bartels, R. H. & Stewart, G. W. Algorithm 432 [C2]: Solution of the matrix equation AX + XB = C [F4]. *Commun. ACM* **15**, 820–826 (1972).

8. Higham, N. J. *Accuracy and Stability of Numerical Algorithms*. (Society for Industrial and Applied Mathematics (SIAM, 3600 Market Street, Floor 6, Philadelphia, PA 19104), Philadelphia, Pa, 2002). doi:10.1137/1.9780898718027.

9. Brask, J. B. Gaussian states and operations -- a quick reference. Preprint at https://doi.org/10.48550/arXiv.2102.05748 (2022).

10. Mahboob, I., Nishiguchi, K., Okamoto, H. & Yamaguchi, H. Phonon-cavity electromechanics. *Nature Phys* **8**, 387–392 (2012).